\documentclass[twocolumn,prl,showpacs,aps]{revtex4-1}

\usepackage[english]{babel}
\usepackage[ansinew]{inputenc}
\usepackage{times}
\usepackage{graphicx}
\usepackage{graphics}
\usepackage{amsmath}
\usepackage{amsfonts}
\usepackage{amssymb}
\usepackage{dsfont}
\usepackage{epstopdf}
\usepackage{makeidx}
\usepackage{subfigure}
\usepackage{color}
\usepackage{pgf}
\usepackage{bm}
\usepackage{tikz} 
\usepackage[normalem]{ulem}

\newcommand{\be}{\begin{equation}}
\newcommand{\ee}{\end{equation}}
\newcommand{\bea}{\begin{eqnarray}}
\newcommand{\eea}{\end{eqnarray}}

\makeindex
\begin{document}
\title{Enhancing Near-Field Heat Transfer in Composite Media: Effects of the Percolation Transition}
\author{W. J. M.  Kort-Kamp}
\affiliation{Instituto de F\'{\i}sica, Universidade Federal do Rio de Janeiro,
Caixa Postal 68528, Rio de Janeiro 21941-972, RJ, Brazil}
\author{P. I. Caneda}
\affiliation{Instituto de F\'{\i}sica, Universidade Federal do Rio de Janeiro,
Caixa Postal 68528, Rio de Janeiro 21941-972, RJ, Brazil}
\author{F. S. S. Rosa}
\affiliation{Instituto de F\'{\i}sica, Universidade Federal do Rio de Janeiro,
Caixa Postal 68528, Rio de Janeiro 21941-972, RJ, Brazil}
\author{F. A. Pinheiro}
\affiliation{Instituto de F\'{\i}sica, Universidade Federal do Rio de Janeiro,
Caixa Postal 68528, Rio de Janeiro 21941-972, RJ, Brazil}
\date{\today}

\begin{abstract}

We investigate the near-field heat transfer between a semi-infinite medium and a nanoparticle made of composite materials. We show that, in the effective medium approximation, the heat transfer can be greatly enhanced by considering composite media, being maximal at the percolation transition. Specifically, for titanium inclusions embedded in a polystyrene sphere, this enhancement can be up to thirty times larger than in the case of the corresponding homogeneous titanium sphere. We demonstrate that our findings are robust against material losses, to changes in the shape of inclusions and materials, and apply for different effective medium theories. These results suggest the use of composite media as a new, versatile material platform to enhance, optimize, and tailor near-field heat transfer in nanostructures.

\end{abstract}
\maketitle
%

Since the seminal work by Polder and van Hove~\cite{PvH1971}, in which it was shown that the near field heat transfer (NFHT) \cite{Greffet2005,Dorofeyev2011,Raschke2013,BiehsIR,Volokitin2007} between two media at short separations can vastly exceed the blackbody limit, numerous works have been carried out to investigate, both theoretically and experimentally, the physics involved in this process. On the theoretical front, the formulation of the NFHT in terms of scattering matrices~\cite{Narayanaswamy2008,Bimonte2009,Messina2011,Kruger2011} opened new venues for investigating the effects of non-trivial geometries, as did the more numerical oriented approaches of fluctuating surface currents \cite{Rodriguez2013} and FDTD computations \cite{Rodriguez2012}. As selected (and by no means exhausting) examples, we can highlight studies of the heat transfer for the sphere-plate configuration \cite{Otey2011}, between gratings \cite{Guerout2012,Lussange2012}, between particles and surfaces \cite{ChapuisEtAl2008,HuthEtAl2010}, tips and surfaces \cite{PBA2013}, and various shapes \cite{Rodriguez2013}. There was also great activity regarding the material properties in the NFHT, like hyperbolic materials~\cite{Biehs2012}, porous media~\cite{BiehsEtAl2011}, photonic crystals~\cite{PBA2010}, and graphene sheets \cite{Volokitin2011}. On the experimental side, several groups carried out measurements of NFHT for different geometries, such as tip-surface~\cite{Kittel}, sphere-plate~\cite{NarayaEtAl2008,RousseauEtAl2009}, and plate-plate~\cite{HuEtAl2008,Ottens2011}, all in fairly good agreement with theoretical predictions. All this development in the field of NFHT has naturally led to investigations of possible applications. Among many ideas, there has been studies in thermal imaging~\cite{BiehsEtAl2008}, thermal rectification and control \cite{Fan2010,Zwol2011,BiehsRosaAPL,BiehsPBA2013}, and optimization of thermophotovoltaic cells \cite{Narayanaswamy2003,Laroche2006}, all of which take advantage of the large increase of the heat flux brought forth by the near field. As a result, enhancing the process of NFHT is crucial for the development of new and/or optimized applications. Indeed, there are some recent proposals in this direction (see {\em e.g.} Refs.~\cite{Guerout2012,Worbes2013}), where enhancements up to a factor of a few tens have been reported.

The aim of this Letter is to introduce a novel approach to enhance the heat transfer in the near field by exploiting the versatile material properties of composite media. To this end we investigate the NFHT between a semi-infinite dielectric medium and metallic nanoparticles, with various concentrations and geometries, embedded in dielectric hosts. Applying the Bruggeman homogenization technique, we demonstrate that the NFHT is strongly enhanced in composite media if compared to the case where homogeneous media are considered. In particular we show that NFHT is maximal precisely at the percolation transition. We also demonstrate that at the percolation transition more modes effectively contribute to the heat flux, widening the transfer frequency band. We show that these results are valid regardless the geometrical shape of the inclusions and are robust against material losses. We hope that our findings might be useful to establish composite media as a novel platform for applications involving NFHT.

\begin{figure}[!h]
  \centering
  \includegraphics[scale = 0.4]{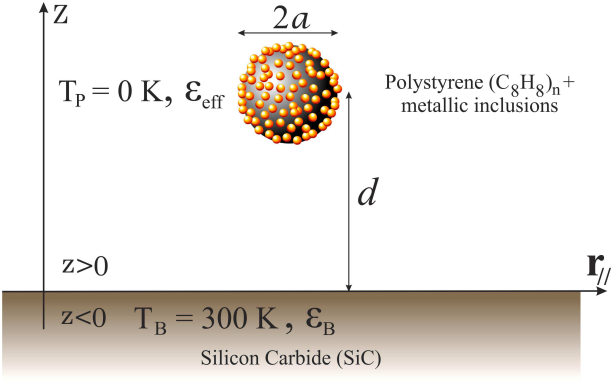}
  \caption{Schematic representation of the system under study.}
  \label{Figure1}
\end{figure}

Let us consider NFHT in the system depicted in Fig. \ref{Figure1}. The half-space $z<0$ is composed of an isotropic and homogeneous (bulk) material (dielectric constant $\varepsilon_B (\omega)$) at temperature $T_B = 300$ K. The upper medium $z>0$ is vacuum and a sphere of radius $a$ at temperature $T_P = 0$ K is located at a distance $d$ above the interface. The spherical particle is made of randomly distributed and oriented metallic spheroids with dielectric function $\varepsilon_{i} (\omega)$, embedded in a host medium with dielectric constant $\varepsilon_{hm}(\omega)$.
Provided the size of the inclusions are much smaller than the relevant wavelengths for NFHT, the effective permittivity $\varepsilon_{e}(\omega, f, L)$ of the composite sphere can be calculated using the well-known Bruggeman effective medium theory (BEMT), which provide a local model for $\varepsilon_{e}(\omega, f, L)$~\cite{Lagarkov1996, Brouers1986,Goncharenko2004, SuppMaterial},
\begin{eqnarray}
(\!\!\!&1&-f)\left\{\dfrac{\varepsilon_{hm} - \varepsilon_{e}}{\varepsilon_{e} + L(\varepsilon_{hm}-\varepsilon_{e})}
+ \dfrac{4(\varepsilon_{hm} - \varepsilon_{e})}{2\varepsilon_{e} + (1-L)(\varepsilon_{hm}-\varepsilon_{e})}\right\}\cr
\!\!\!&+&f\!\!\left\{\dfrac{\varepsilon_{i} - \varepsilon_{e}}{\varepsilon_{e} + L(\varepsilon_{i}-\varepsilon_{e})}
+ \dfrac{4(\varepsilon_{i} - \varepsilon_{e})}{2\varepsilon_{e} + (1-L)(\varepsilon_{i}-\varepsilon_{e})}\right\}=0 \, \label{BEMT} ,
\end{eqnarray}
where $0\leq f \leq 1$ and $0 \leq L \leq 1$ are the filling and depolarization factors of inclusions, respectively. It is worth mentioning that other homogenization techniques and mixing rules do exist, but the BEMT distinguishes itself for being the simplest analytical model that predicts an insulator-metal transition at a nontrivial filling factor~\cite{Lagarkov1996, Brouers1986,Goncharenko2004,Choy1999, Sahimi1993}.

The NFHT process is governed by fluctuating currents in the bulk and the composite particle. The currents in the bulk, in local thermal equilibrium, induce electromagnetic fields that eventually illuminate the particle. If the relevant wavelengths to the NFHT are much larger than $a$, and $d$ is of order of a few radii, the electromagnetic response of the sphere can be described in terms of its electric and magnetic dipoles~\cite{ChapuisEtAl2008,HuthEtAl2010,magneticdipole}. Here we do not take into account the diamagnetic response of the material so that the magnetic dipole moment is due to eddy currents in the composite particle. In this case, the mean power per unit of frequency radiated by the bulk and absorbed by the composite particle can be cast as~\cite{ChapuisEtAl2008,HuthEtAl2010}
\begin{eqnarray}
{\cal {P}}_{\textrm{abs}}(\omega, f, L, d)&=& \omega\textrm{Im}[\alpha_E(\omega, f, L)] \varepsilon_0\langle |{\bf E}|^2 \rangle \cr\cr
&+&    \omega \textrm{Im}[\alpha_H(\omega, f, L)] \mu_0\langle |{\bf H}|^2 \rangle\, ,
\label{pabs}
\end{eqnarray}
where ${\bf E}$ and ${\bf H}$ are the electric and magnetic fields impinging on the particle and $\langle ... \rangle$ denotes statistical average over bulk current fluctuations. Also $\alpha_E(\omega, f, L)$ and $\alpha_H(\omega, f, L)$ are the electric and magnetic polarizabilities of the composite particle, calculated via Mie scattering theory~\cite{ChapuisEtAl2008,HuthEtAl2010,magneticdipole,SuppMaterial}.

In the following calculations we take the semi-infinite medium to be made of silicon carbide (SiC) and a composite medium of randomly dispersed spheroidal copper in a host sphere of polystyrene $(C_8H_8)_n$. The dispersive models for these materials are well known and were taken from the references~\cite{Greffet2005, Ordal1985, Hough80, SuppMaterial}. The sphere's radius is $a = 50$ nm and the distance between the particle and the half-space is $d = 200$ nm. We have verified that for the materials and geometric parameters chosen the applicability of the dipole approximation is guaranteed, and contributions from higher multipoles and multiple scattering, which are not taken into account in Eq.~(\ref{pabs}), are negligible to NFHT.
\begin{figure}[!h]
  \centering
  \includegraphics[scale = 0.4]{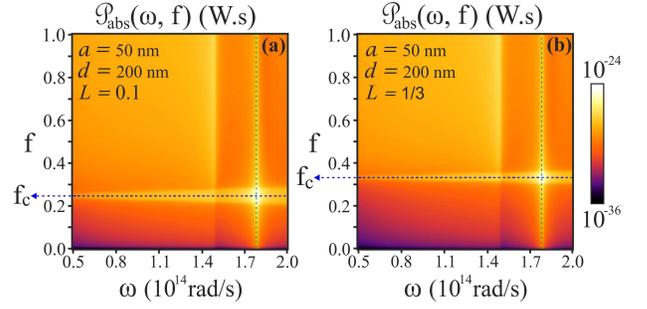}\\
  \caption{Mean power absorbed by a polysterene particle with embedded copper inclusions as a function of frequency and filling factor for a fixed distance between the nanoparticle and the SiC medium and two different values of the depolarization factor, {\bf (a)} $L = 0.1$ and {\bf (b)} $L=1/3$. In both cases, the horizontal dashed lines correspond to percolation threshold $f_{c}$ predicted by the Bruggeman effective medium theory whereas the vertical dashed lines correspond to the position of plasmon resonance for SiC.}
  \label{Figure2}
\end{figure}

In Fig.~\ref{Figure2} the mean power absorbed by the particle ${\cal {P}}_{\textrm{abs}}(\omega, f, L, d)$ between $\omega$ and  $\omega + d \omega $ is calculated as a function of frequency and the volume fraction $f$ for two different values of the depolarization factor $L$, which encodes all the information related to the microgeometry of the inclusions: $L=0.1$ (needle-like particles) and $L=1/3$ (spherical inclusions). In both cases, there is a strong enhancement in ${\cal {P}}_{\textrm{abs}}$ that are related to the excitation of surface phonon polaritons in the bulk that occur for $\textrm{Re} \left[{\varepsilon}_{B}(\omega_{P})\right] = -1$, related to a peak in the density of states at $\omega_{P}$~\cite{Greffet2005}. For SiC, $\omega_{P} \approx 1.787 \times10^{14}$ rad/s \cite{Greffet2005}, as shown by the vertical dashed lines.
Also, it is clear from Fig.~\ref{Figure2} that there exists a value of the volume fraction $f$ for which the absorbed power by the particle is maximal. For $L=0.1$ this peak occurs at $f_{m} \approx 0.25$ whereas for $L=1/3$ it shows up for $f_{m} = 1/3$. It is also important to emphasize that: {\it (i)} there is a broadening of the spectral heat flux at  $f_{m}$, {\it i.e.} more modes effectively contribute to the NFHT process; {\it (ii)} for any frequency the maximal enhancement in ${\cal {P}}_{\textrm{abs}}$ occurs at $f_{m}$, as it can be seen from Fig.~\ref{Figure2} for both spheres and needle-like particles. Remarkably, these values of $f_{m}$ correspond exactly to the percolation threshold $f_{c}$ predicted by the BEMT~\cite{Brouers1986,Lagarkov1996,Goncharenko2004}
\begin{equation}
f_{c}(L) = \frac{L(5-3L)}{(1+ 9L)}.
\label{bruggeman}
\end{equation}
The percolation threshold $f_c$ corresponds to a critical value in the filling factor for which the composite media undergoes a insulator-conductor transition and the system exhibits a dramatic change in its electrical and optical properties~\cite{Goncharenko2004,Brouers1986,Lagarkov1996,Sarychev2000,SuppMaterial}.

\begin{figure}[!h]
  \centering
  \includegraphics[scale = 0.4]{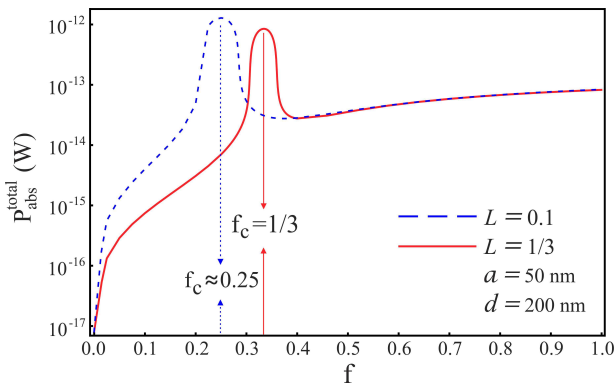}\\
 \caption{Total power absorbed by the composite particle as a function of $f$ for $L = 0.1$ (blue dashed line) and $L=1/3$ (solid red line). The vertical arrows highlight that the values of maximum heat transfer occur precisely at the percolation threshold $f_{c}$ given by Eq.~(\ref{bruggeman}). The other numerical parameters are the same as in Fig.~\ref{Figure2}.}
  \label{Figure3}
\end{figure}

To further investigate the effects of percolation on the NFHT process, in Fig.~\ref{Figure3} we depict the total power absorbed by the composite particle $P^{total}_{abs} (f, L, d)$, calculated as function of the volume fraction $f$ for $L=0.1$ and $L=1/3$. From Fig.~\ref{Figure3} it is clear that $P^{total}_{abs}$ is maximal at the percolation threshold $f_{c}$ for the two inclusions geometries, confirming that NFHT is greatly enhanced at the percolation critical point. It also very important to stress that the simple fact of considering a composite particle, even for inclusion concentration far from $f_{c}$, often enhances the NFHT process if compared to the case where the materials involved are homogeneous. Indeed, the enhancement factor in $P^{total}_{abs}$ due the inclusion of copper nanoparticles can be as high as $15$ if compared to the case of an homogeneous particle made of copper ($f=1$) and $10^{5}$ if compared to the one of a homogeneous polystyrene particle ($f=0$). This result unambiguously demonstrates that composite media can largely outperform homogeneous media in NFHT, which therefore may find novel applications and optimize heat transfer at the nanoscale.

\begin{figure}[!h]
  \centering
  \includegraphics[scale = 0.4]{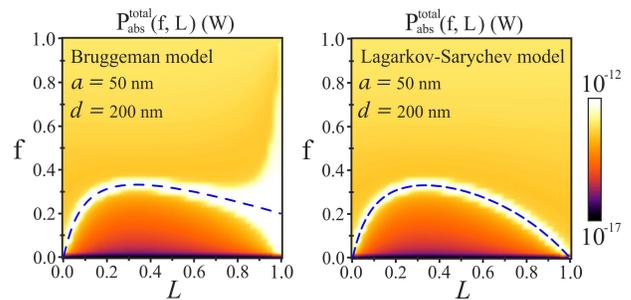}\\
  \caption{Contour plot of $P^{total}_{abs}$ as a function of both the depolarization factor $L$ and filling factor $f$. The dashed blue curve corresponds to the critical filling factor $f_{c}$ that determines the percolation threshold for (a) Bruggeman effective medium theory and (b) Lagarkov-Sarychev model. The other numerical parameters are the same as in Fig.~\ref{Figure2}.}
  \label{Figure4}
\end{figure}

The dependence of $P^{total}_{abs}$ on the shape of the copper inclusions is investigated in Fig.~\ref{Figure4}a, where $P^{total}_{abs} (f,L,d)$ is calculated as a function of both $f$ and the depolarization factor $L$, which only depends on the geometry of the inclusions~\cite{Goncharenko2004,Brouers1986,Lagarkov1996}. Figure~\ref{Figure4}a reveals that the maximal enhancement in $P^{total}_{abs} (f,L,d)$ occurs at $f_{c}$ not only for the two particular inclusion geometries considered above (spheres and needle-like particles) but for all possible spheroids. Indeed, the value of the filling factor that leads to maximal $P^{total}_{abs} (f,L,d)$ corresponds precisely to the prediction of the percolation threshold $f_{c}$ of the BEMT [Eq.~(\ref{bruggeman})] for all $L$, demonstrating the robustness of our findings against the variation of the shape of the inclusions. For $L \gtrsim 0.7$ (oblate spheroids) the global maximum in $P^{total}_{abs} (f,L,d)$ becomes more broadly distributed around the percolation threshold; nevertheless, on average, it still occurs at $f_{c}$. For copper inclusions, the ratio between the total power absorbed by the particle at the percolation threshold, $P^{total}_{abs} (f=f_{c},L,d)$, and its value for an homogeneous sphere made of the same material of the inclusions, $P^{total}_{abs} (f=1,L,d)$, is 15.4 for $L=0.1$ (needle-like inclusions) and 10.4 for $L=1/3$ (spherical inclusions). For titanium inclusions this ratio is even larger; it can be as high as 28.7 for $L=0.1$. The values of the ratio $P^{total}_{abs} (f=f_{c},L,d)/P^{total}_{abs} (f=1,L,d)$ for several metals is shown in Table I. For all investigated metallic materials the enhancement in NFHT is maximal at $f_{c}$, for every $L$, a fact that suggests that our findings are independent of the metals of choice.

In order to test the robustness of our results against modifications of the effective medium theory, in Fig.~\ref{Figure4}b we depict, as a function of $L$ and $f$, the total power $P^{total}_{abs}$ absorbed by an inhomogeneous particle with its effective electric permittivity being obtained by means of an alternative homogenization technique, namely the one proposed in Ref.~\cite{Lagarkov1996}. That effective electric permittivity (explicitly written in the supplemental material \cite{SuppMaterial}) is known to give more accurate results for $f_{c}$ than the BEMT in the regime of small $L$ ($L \ll 1 $); it predicts that $f_{c}$ follows from Eq. (7) of~\cite{SuppMaterial}, which is different from the prediction (\ref{bruggeman}) of the BEMT. Figure~\ref{Figure4}b reveals that, using this alternative effective medium prescription, the maximal value for $P^{total}_{abs}$ again occurs at the percolation threshold for all $L$, as it also happens within the BEMT. This fact suggests that the maximal enhancement of the NFHT in inhomogeneous media at the percolation threshold is, at least to a certain degree, independent of the effective medium theory utilized.

In order to understand the physical mechanism leading to the strong enhancement of the NFHT at the percolation threshold $f_{c}$, we recall that the metal-insulator transition associated with percolation is a geometric phase transition where current and electric field fluctuations are expected to be large and scale invariant~\cite{Sarychev2000}. In composite mixtures of metallic grains embedded in dielectric hosts, these strong fluctuations at $f_{c}$ induce a local electric field concentration (``hot spots")~\cite{Losquin2013,Caze2013} at the edge of metal clusters; the distance between field maxima is of the order of the correlation percolation length~\cite{Sarychev2000,Stauffer1994}. In addition, for an ideally loss-free ($\textrm{Im} [\varepsilon_{hm}] = \textrm{Im} [\varepsilon_{i}]=0$) inhomogeneous mixture of metallic grains embedded in a dielectric, at the percolation critical point the effective electric permittivity $\varepsilon_{e}$ is mainly imaginary ($\textrm{Im} [\varepsilon_{e}] \gg \textrm{Re} [\varepsilon_{e}$]) so that the composite medium is highly absorptive~\cite{Gadenne1998}.  Hence the electric fields localized at the ``hot spots", and consequently the electromagnetic energy stored in medium, are expected to increase unlimitedly at $f_{c}$. These local fields of course remain finite due to unavoidable losses but are still very large at $f_{c}$, resulting in maximal absorption by the composite medium and explaining why there is a peak in $P^{total}_{abs}$  precisely at the percolation threshold. Furthermore in our particular system, where we have considered realistic material losses, we have verified that the relation $\textrm{Im} [\varepsilon_{e}] \gg \textrm{Re} [\varepsilon_{e}]$ still holds at $f_{c}$. Nonlinear optical phenomena are expected to be important at the percolation threshold due the enhancement of the local electric field~\cite{Sarychev2000,Gadenne1998}. Despite the fact that we have neglected nonlinearities in the present study for the sake of simplicity, we anticipate that they should lead to an even more significative enhancement of $P^{total}_{abs}$ at $f_{c}$. Finally, it is important to emphasize that the above arguments to explain the enhancement of NFHT at $f_{c}$ rely on important critical properties of the percolation phase transition, which are known to be independent of the details of the effective medium model of choice~\cite{Sarychev2000}. This reasoning, together with the fact that our results are found to be independent of the investigated homogenization techniques, provide evidence that our findings should hold even beyond the effective medium approximation. 
\begin{table}
\centering
\begin{tabular}{|c|c|c|}
\hline
 & $L=0.1$ & $L=1/3$ \\ \hline
Titanium & 28.7 & 20.7 \\ \hline
Copper & 15.4 & 10.4 \\ \hline
Vanadium  & 7.0 & 5.0\\ \hline
Silver & 5.1 & 3.7 \\ \hline
Gold & 3.6 & 2.6 \\ \hline
\end{tabular}
\label{Tab1}
\caption{Ratio between the total power absorbed by the inhomogeneous particle at the percolation threshold, $P^{total}_{abs} (f=f_{c},L,d)$, and its value for an homogeneous sphere made of the same metal of the inclusions, $P^{total}_{abs} (f=1,d)$, for several metals and for $L=0.1$ (needle-like inclusions) and 10.4 for $L=1/3$ (spherical inclusions).}
\end{table}

In conclusion, we have investigated the near-field heat transfer between a half-space and a nanoparticle made of composite materials. For concreteness, we have considered realistic materials usually employed in experiments of near-field heat transfer at the nanoscale: a polystyrene sphere with embedded metallic occlusions and a SiC semi-infinite medium. Using the Bruggeman effective medium theory, we show that heat transfer between the nanoparticle and the half-space is largely enhanced by the fact that the particle contains randomly distributed inclusions; the enhancement factor can be as large as thirty if one compares to the case of an homogeneous metallic sphere. We also demonstrate that heat transfer is maximal at the percolation threshold in the nanoparticle for all possible spheroids, a result we show to be robust against material losses. We argue that this effect is related to the critical properties of the percolation phase transition, such as enhanced fluctuations of currents and electromagnetic fields inside the particle, which are known to be universal and independent of the details of the effective medium model. Our findings suggest that composite media can be used as a new, versatile material platform, of easy fabrication, to tailor and optimize near-field heat transfer at the nanoscale.

We thank S.L.A de Queiroz and C. Farina for useful discussions, and FAPERJ, CNPq, and CAPES for financial support. One of us (P.I.C.) acknowledges PIBIC/UFRJ for financial support.


\end{document}


\begin{center}
{\Large {\bf Supplemental material to ``Enhancing Near-Field Heat Transfer in Composite Media: Effects of the Percolation Transition"}}\\ \vspace{0.5cm}
{\large W. J. M.  Kort-Kamp,
%
P. I. Caneda,
%
F. S. S. Rosa,
%
F. A. Pinheiro}\\ \vspace{0.2cm}
Instituto de F\'{\i}sica, Universidade Federal do Rio de Janeiro,\\
Caixa Postal 68528, Rio de Janeiro 21941-972, RJ, Brazil
\end{center}
%
\section{Effective medium theories}

Effective medium theories (EMT) allow one to determine, by means of algebraic formulas, the effective electric permittivity $\varepsilon_e$ of a composite medium as a function of the constituent permittivities and shapes as well as of the fractional volumes characterizing the mixture [36-38].
One of the most important and successful EMTs is the Bruggeman Effective Medium Theory (BEMT), which is the simplest analytical model that predicts a metal-insulator transition at a critical volume fraction, $f_{c}$. BEMT treats the dielectric host medium and the metallic inclusions symmetrically, and it is based on the following assumptions:
$(i)$ the grains are assumed to be randomly oriented spheroidal particles, and $(ii)$ they are embedded in an homogeneous effective medium that will be determined self-consistently. If a quasi-static electromagnetic field ${\bf E}_0$ impinges on such an inhomogeneous medium, the electric field ${\bf E}_i^{in}$ inside the metallic inclusions (permittivity $\varepsilon_i$), and the field ${\bf E}_{hm}^{in}$ inside the dielectric grains (permittivity $\varepsilon_{hm}$) read [37]
\begin{eqnarray}
{\bf E}_i^{in} &=& \left[\dfrac{1}{3} \dfrac{\varepsilon_e}{\varepsilon_e + L(\varepsilon_i - \varepsilon_e)}
+ \dfrac{2}{3}  \dfrac{2\varepsilon_e}{2\varepsilon_e + (1-L)(\varepsilon_i - \varepsilon_e)}\right]{\bf E}_0\, ,\\
{\bf E}_{hm}^{in} &=& \left[\dfrac{1}{3} \dfrac{\varepsilon_e}{\varepsilon_e + L(\varepsilon_{hm} - \varepsilon_e)}
+ \dfrac{2}{3}  \dfrac{2\varepsilon_e}{2\varepsilon_e + (1-L)(\varepsilon_{hm} - \varepsilon_e)}\right]{\bf E}_0\, ,
\end{eqnarray}
where $0\leq L \leq 1$ is the depolarization factor of the spheroidal inclusions. For an arbitrary ellipsoidal particle characterized by the semi-axes $a_x\, , \ a_y\, ,\ a_z$,  the depolarization factor $L_i$ in the $i-$direction is [17]
\begin{eqnarray}
L_i = \dfrac{a_xa_ya_z}{2} \int_0^{\infty} \dfrac{ds}{(s+a_i^2)\sqrt{(s+a_x^2)(s+a_y^2)(s+a_z^2)}}\, ,
\end{eqnarray}
where the relation $L_x+L_y+L_z=1$ holds. In this work we consider spheroidal inclusions such that $L_z = L$ and $L_x = L_y = (1-L)/2$.

Hence, the effective permittivity $\varepsilon_{e}$ is defined by
\begin{eqnarray}
\langle{\bf D} \rangle = \varepsilon_e \langle {\bf E}\rangle \rightarrow
 f\varepsilon_i {\bf E}_i^{in} + (1-f) \varepsilon_{hm}{\bf E}_{hm}^{in} = \varepsilon_e f {\bf E}_i^{in} + \varepsilon_{e} (1-f) {\bf E}_{hm}^{in} \, ,
 \label{epsilonefetivo}
\end{eqnarray}
where $0\leq f \leq 1$ is the volume filling factor for the inclusions.
Equation (\ref{epsilonefetivo}) implies that $\varepsilon_{e}$ satisfies exactly Eq. (1) given in the paper, which is valid provided the typical size of the grains is much smaller than the incident wavelength.
%
%
Also, it is worth mentioning that the self-consistent equation describing $\varepsilon_e$ [Eq. (1) in the paper] has several roots, but only the one with $\textrm{Im}(\varepsilon_e) \geq 0$ is physical. 

Finally, the percolation threshold $f_{c}$ is calculated by taking the limit $\omega \rightarrow 0$, and evaluating the effective conductivity and permittivity of the composite medium. In the case of BEMT, Eq.~(1) in the paper leads to the following expression for the percolation threshold [36, 37, 38]
\begin{equation}
f_{c}(L) = \frac{L(5-3L)}{(1+ 9L)}.
\label{bruggeman}
\end{equation}

In addition to BEMT, there are many different versions of effective medium theories; an important example is the one developed by Lagarkov-Sarychev [37]. It is known to predict more accurate results for $f_{c}$ than the BEMT in the case of very elongated inclusions (needle-like inclusions, small $L$). Within this approach $\varepsilon_{e}$ is obtained following a self-consistent prescription that is analogous to the one of BEMT, but under the assumption that the dielectric and metallic grains should not be treated symmetrically. Rather, the host medium is supposed to have spherical symmetry, whereas the metallic inclusions are considered to be spheroidal particles; this approach leads to [37]:
\begin{eqnarray}
9(1-f)\left\{\dfrac{\varepsilon_{hm} - \varepsilon_{e}}{2\varepsilon_{e} + \varepsilon_{hm}}\right\} +
f\left\{\dfrac{\varepsilon_{i} - \varepsilon_{e}}{\varepsilon_{e} + L(\varepsilon_{i}-\varepsilon_{e})}
+ \dfrac{4(\varepsilon_{i} - \varepsilon_{e})}{2\varepsilon_{e} + (1-L)(\varepsilon_{i}-\varepsilon_{e})}\right\}=0\, .
\end{eqnarray}
The percolation threshold $f_{c}$ predicted by the Lagarkov-Sarychev effective medium theory is [37]
\begin{eqnarray}
f_c(L) = \dfrac{9L(1-L)}{2+15L-9L^2}\, .
\end{eqnarray}

\section{Electric and Magnetic Polarizabilities of Small Particles}
The sphere's electric, $\alpha_E$, and  magnetic, $\alpha_H$, polarizabilities are given by [16, 17, 43]
\begin{eqnarray}
\dfrac{\alpha_E(\omega, f, L)}{(4\pi a^3/3)} &=& \dfrac{3}{2} \dfrac{(2y^2+x^2)[\sin{y}-y\cos{y}]-x^2y^2\sin{y}}
{(y^2-x^2)[\sin{y}-y\cos{y}]-x^2y^2\sin{y}}\, , \cr \cr
\dfrac{\alpha_H(\omega, f, L)}{(4\pi a^3/3)} &=& \dfrac{1}{4} \left[\dfrac{(x^2-6)}{y^2}(y^2+3y\cot{y}-3)-\dfrac{2x^2}{5} \right],
\end{eqnarray}
where $x = \omega a/c$ and $y = \sqrt{\varepsilon_{e}(\omega, f, L)}x$.

\begin{figure}[!h]
  \centering
  \includegraphics[scale = 0.7]{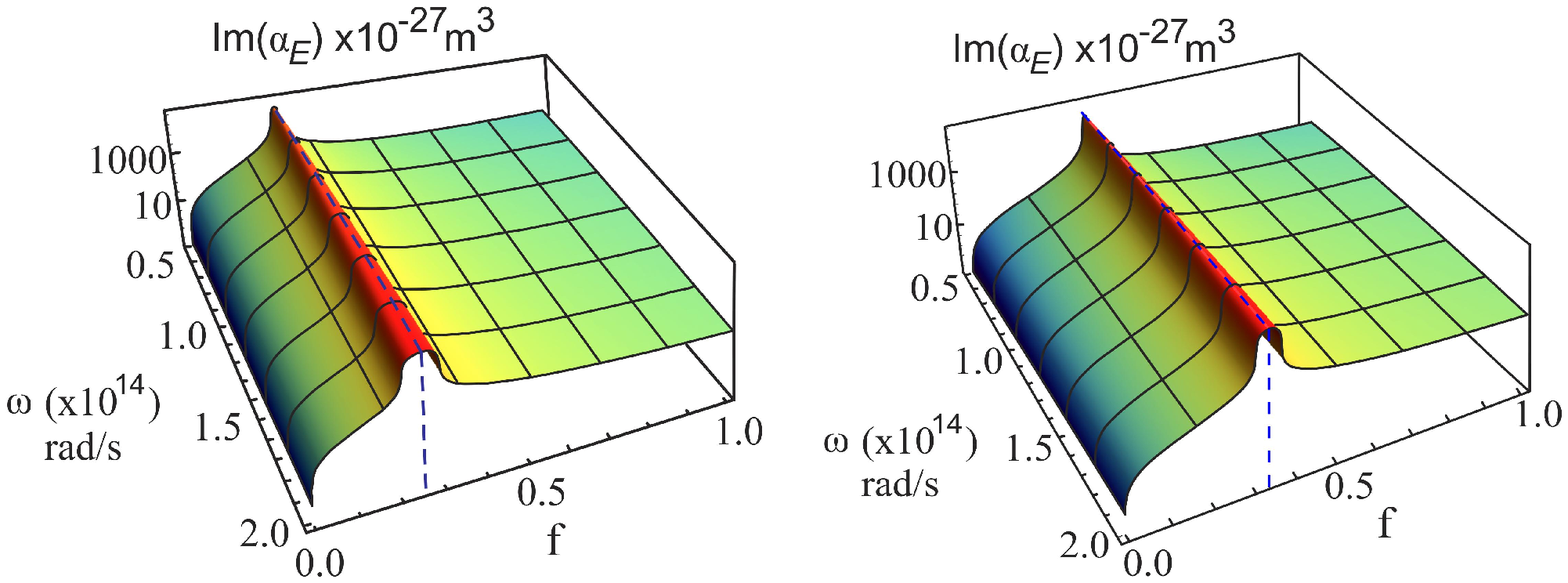}
  \caption{Electric polarizability of a inhomogeneous sphere made of copper inclusions randomly distributed in a polystyrene spherical host as a function of frequency $\omega$ and volume fraction $f$ for two different values of the depolarization factor $L$: (a) $L=0.1$ and (b) $L=1/3$.}
  \label{Figure1}
\end{figure}

In Fig.~\ref{Figure1} $\textrm{Im} (\alpha_E)$ is calculated, within the BEMT, as a function of both frequency $\omega$ and filling factor $f$ for two different values of $L$: $(a) L=0.1$ and $(b) L=1/3$, for an inhomogeneous composed of copper inclusions randomly distributed inside a polystyrene host sphere. Fig.~\ref{Figure1} reveals that, for every frequency, the maximal value of $\textrm{Im} (\alpha_E)$ occurs at $f_{c}$. This result corroborates the fact that the maximal power absorbed by the particle in the NFHT is maximal at the percolation threshold. We have verified that this result applies for every value of $L$ (only two distinctive examples are shown in Fig.~\ref{Figure1}). Figure~\ref{Figure1} also shows that, for a given $f$, $\textrm{Im} (\alpha_E)$ weakly depends on frequency, as it remains almost constant as one varies $\omega$.

\begin{figure}[!h]
  \centering
  \includegraphics[scale = 0.7]{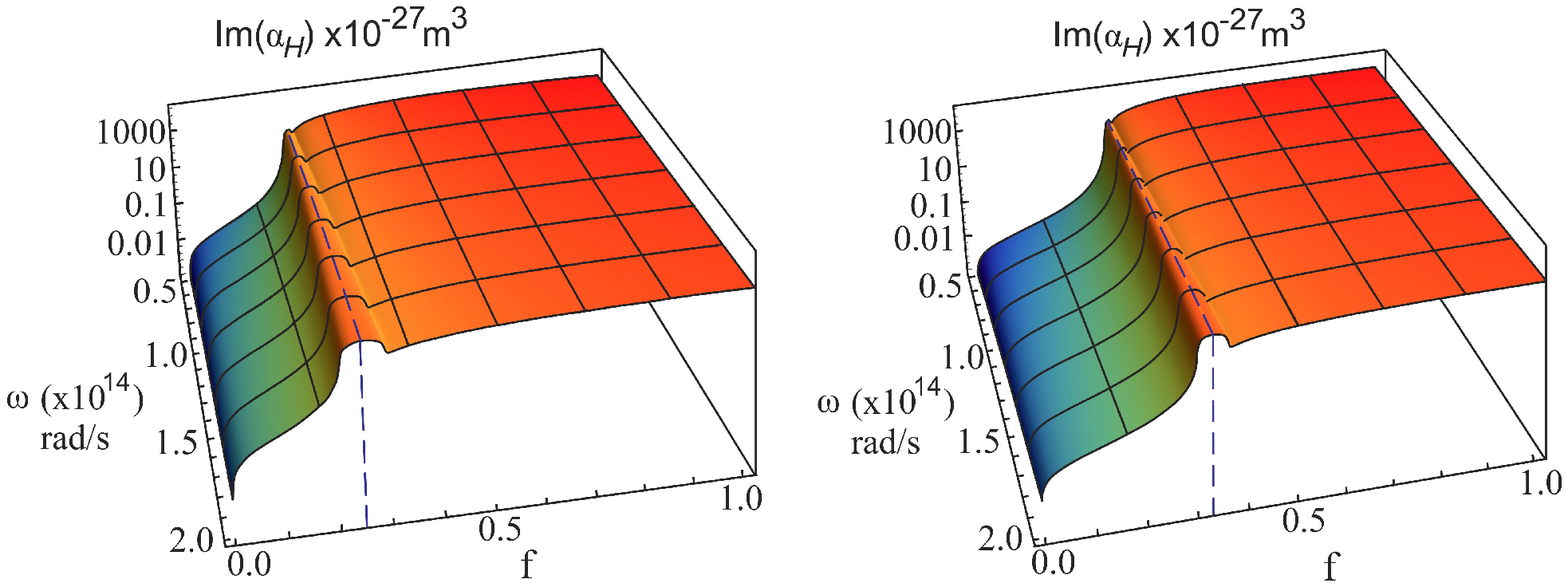}
   \caption{Magnetic polarizability of a inhomogeneous sphere made of copper inclusions randomly distributed in a polystyrene spherical host as a function of frequency $\omega$ and volume fraction $f$ for two different values of the depolarization factor $L$: (a) $L=0.1$ and (a) $L=1/3$.}
  \label{Figure2}
\end{figure}

In Fig.~\ref{Figure2} $\textrm{Im} (\alpha_H)$ is shown as a function of both frequency $\omega$ and filling factor $f$ for the same parameters of Fig.~\ref{Figure1}. As it occurs for $\textrm{Im} (\alpha_E)$, $\textrm{Im} (\alpha_H)$ is weakly dependent on frequency for fixed $f$. However, $\textrm{Im} (\alpha_H)$ is much less peaked around $f_{c}$ than $\textrm{Im} (\alpha_E$), although a local maximum in $\textrm{Im} (\alpha_H)$ always exists at $f_{c}$ for every $L$. The joint contribution of the maxima at $f_{c}$, which occurs for both $\textrm{Im} (\alpha_E)$ and $\textrm{Im} (\alpha_H)$, ultimately leads to the large enhancement of NFHT for composite media at the percolation critical point.

\section{Material parameters}
The electrical permittivity $\varepsilon_{B}$ of Silicon Carbide (SiC) is well described by following dispersive model [2]
\begin{equation}
\varepsilon_{B}(\omega) = \epsilon_{\infty}\left(1 + \dfrac{\omega_L^2 - \omega_T^2}{\omega_T^2 - \omega^2 - i\Gamma \omega}\right),
\label{sic}
\end{equation}
where $\omega_L = 182.7\times10^{12}$ rad/s, $\omega_T = 149.5\times10^{12}$ rad/s, and $\Gamma = 0.9\times10^{12}$ rad/s.

For the polystyrene host medium, its electric permittivity $\varepsilon_{hm}$ reads [45]
\begin{equation}
\varepsilon_{hm}(\omega) = 1 + \dfrac{\omega_{p1}^2}{\omega_{r1}^2 - \omega^2 - i\Gamma_{1}\omega} +\dfrac{\omega_{p2}^2}{\omega_{r2}^2 - \omega^2 - i\Gamma_{2}\omega},
\end{equation}
where $\omega_{p1} = 1.11\times10^{14}$ rad/s, $\omega_{r1} = 5.54\times10^{14}$ rad/s, $\omega_{p2} = 1.96\times10^{16}$ rad/s, $\omega_{r2} = 1.35\times10^{16}$ rad/s, and  $\Gamma_1 = \Gamma_2 = 0.1 \times 10^{12}$ rad/s.

The metallic inclusions are described by the Drude model [44]
\begin{equation}
\varepsilon_{i}(\omega) = 1-\dfrac{\omega_{i}^2}{\omega^2+ i \Gamma_{i} \omega},
\end{equation}
where the material parameters for metallic inclusions of Titanium (Ti), Copper (Cu), Vanadium (V), Silver (Ag), and Gold (Au) are given in Table I.
\begin{table}[h!]
\centering
\begin{tabular}{|c|c|c|}
\hline
 & $\omega_i$ (rad/s) & $\Gamma_i$ (rad/s) \\ \hline
Titanium &  $3.83 \times 10^{15}$  & $7.19 \times 10^{13}$  \\ \hline
Copper & $1.12 \times 10^{16}$ & $1.38 \times 10^{13}$ \\ \hline
Vanadium   & $7.84 \times 10^{15}$ & $9.26 \times 10^{13}$\\ \hline
Silver & $1.37 \times 10^{16}$ & $2.73 \times 10^{13}$ \\ \hline
Gold & $1.37 \times 10^{16}$  & $4.05 \times 10^{13}$ \\ \hline
\end{tabular}
\label{Tab1}
\caption{Material parameters used in the dispersive Drude model for metallic inclusions.}
\end{table}